\documentclass{article}
\usepackage{spconf,amsmath,graphicx}
\usepackage{subcaption}
\usepackage{stfloats}
\usepackage{multirow}
\usepackage[raggedrightboxes]{ragged2e}
\usepackage{amsfonts} 
\usepackage{amsmath}%
\usepackage{MnSymbol}%
\usepackage{wasysym}%

\usepackage{xcolor}
\usepackage{caption}
\DeclareCaptionJustification{justified}{\justifying}
\title{OMRA: Online Motion Resolution Adaptation to Remedy Domain Shift in Learned Hierarchical B-frame Coding}
\DeclareRobustCommand*{\IEEEauthorrefmark}[1]{%
\raisebox{0pt}[0pt][0pt]{\textsuperscript{\footnotesize\ensuremath{#1}}}}

\name{Zong-Lin Gao\IEEEauthorrefmark{1}, Sang NguyenQuang\IEEEauthorrefmark{1}, Wen-Hsiao Peng\IEEEauthorrefmark{1}, and Xiem HoangVan\IEEEauthorrefmark{2} }
\address{  \IEEEauthorrefmark{1}Computer Science Department, National Yang Ming Chiao Tung University, Taiwan \\
\IEEEauthorrefmark{2}Electronics and Telecommunications, VNU - University of Engineering and Technology, Vietnam}
\begin{document}
%
\maketitle
\begin{abstract}
Learned hierarchical B-frame coding aims to leverage bi-directional reference frames for better coding efficiency. However, the domain shift between training and test scenarios due to dataset limitations poses a challenge. This issue arises from training the codec with small groups of pictures (GOP) but testing it on large GOPs. Specifically, the motion estimation network, when trained on small GOPs, is unable to handle large motion at test time, incurring a negative impact on compression performance. To mitigate the domain shift, we present an online motion resolution adaptation (OMRA) method. It adapts the spatial resolution of video frames on a per-frame basis to suit the capability of the motion estimation network in a pre-trained B-frame codec. Our OMRA is an online, inference technique. It need not re-train the codec and is readily applicable to existing B-frame codecs that adopt hierarchical bi-directional prediction. Experimental results show that OMRA significantly enhances the compression performance of two state-of-the-art learned B-frame codecs on commonly used datasets. 

\end{abstract}


\vspace{-1.2em}
\section{Introduction}
\label{sec:intro}

Learned video compression~\cite{dvclu,ssf,nvc,fvc} has recently shown very promising compression performance. However, most research focuses on P-frame coding. In contrast, learned B-frame coding~\cite{Wu_2018, hlvc, murat_lhbdc, BEPIC, bcanf, maskCRT}, which leverages both the past and future reference frames for improved coding efficiency, is still in its infancy stage. 

There are only a few prior works on learned B-frame coding. Wu~\emph{et al.}~\cite{Wu_2018} presents an early attempt at B-frame coding. It extracts multi-scale features from the future and past reference frames to condition the coding of a target B-frame. Other works~\cite{hlvc, murat_lhbdc, BEPIC, bcanf, TLZMC} encode B-frames following a hierarchical prediction structure. To save motion overhead, Yang~\emph{et al.}~\cite{hlvc} derives motion for multiple reference frames from a single optical flow map. Pourreza~\emph{et al.}~\cite{BEPIC} re-uses a P-frame codec for B-frame coding by first interpolating between the two reference frames to formulate a temporal predictor, followed by coding the target B-frame in a way the same as coding a P-frame. Yılmaz~\emph{et al.}~\cite{murat_lhbdc} adopts the traditional residue-based B-frame coding framework, replacing most of the components with deep networks. Chen~\emph{et al.}~\cite{bcanf} introduces conditional augmented normalizing flows (CANF) for conditional B-frame coding, a scheme known as B-CANF. Notably, B-CANF supports low-delay B-frame coding, where both reference frames are from the same past decoded frame. Alexandre \emph{et al.} \cite{TLZMC} extends B-CANF with a two-layer CANF architecture to skip motion coding. More recently, Chen~\emph{et al.}~\cite{maskCRT} introduces a mask conditional residual Transformer (termed MaskCRT) for learned video coding, supporting both P-frame and B-frame coding.


\begin{figure}[t]
    \centering
    \centerline{\includegraphics[width=0.5\textwidth]{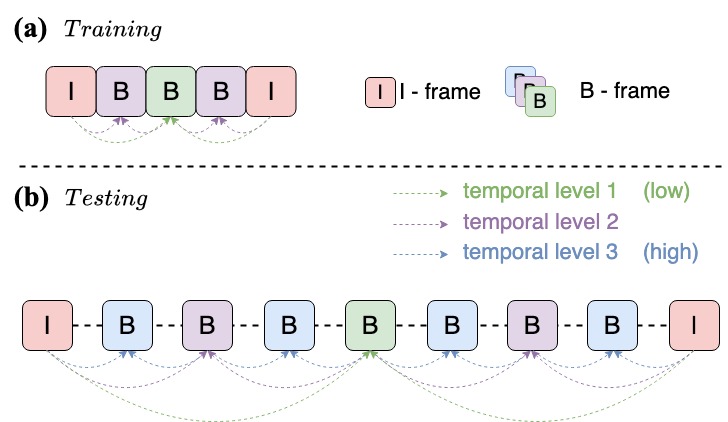}}
    \caption{Illustration of the hierarchical B-frame prediction structures: (a) a 5-frame GOP for training and (b) a 32-frame GOP for test.}
    \label{fig:coding_structure}
    \vspace{-1.2em}
\end{figure}

Despite the recent advances in learned B-frame coding, the state-of-the-art learned B-frame codecs generally show inferior compression performance to their P-frame counterparts. Theoretically, B-frame coding with multi-hypothesis prediction has great potential to outperform P-frame coding with single-hypothesis prediction. This is not the case with learned codecs. As noted in~\cite{bcanf}, one root cause is the domain shift issue, which arises from training learned B-frame codecs on short video sequences with a small group of pictures (GOP) but putting them to use on a large GOP. Most learned codecs, B-frame or P-frame, are trained on Vimeo-90k~\cite{vimeo}, where each training video is only 7-frame long. Fig.~\ref{fig:coding_structure}(a) depicts a common 5-frame GOP structure for training B-frame codecs. At test time, a much larger GOP, as shown in Fig.~\ref{fig:coding_structure}(b), is used. The motion estimation network trained and optimized for small GOPs has difficulties in handling large motion often present in large GOPs. 
As a testimony, Fig.~\ref{fig:ba_visualization}(h) shows the temporal predictor for a video frame right in the middle of a GOP. That is, the motion-compensated frame generated from the two farthest reference frames. The long prediction distance at test time results in a poor-quality warped frame due to unreliable motion estimates, as evidenced by the noticeable differences in comparison with the ground truth in Fig~\ref{fig:ba_visualization}(b).

In this work, we present an online motion resolution adaption (OMRA) method to remedy the domain shift issue by adapting the motion resolution at test time on a per-frame basis without re-training the codec. Our approach is motivated by two key facts: (1) creating a new dataset with a variety of long training sequences covering a wide range of motion characteristics is a non-trivial task, and (2) training B-frame codecs on long training sequences is time-consuming. Recognizing that the motion estimation network trained on small GOPs may not handle well large motion inherent in large GOPs, we downsample both the target frame and its reference frames before conducting motion estimation in order to suit the capability of the motion estimation network. Notably, a number of downsampling factors is evaluated based on minimizing the rate-distortion cost. This is followed by signaling the low-resolution flow maps in the bitstream and super-resolving them for motion compensation in the original resolution. 

Although conceptually simple, our approach is readily applicable to learned B-frame codecs. We test its performance on two state-of-the-art codecs, B-CANF~\cite{bcanf} and MaskCRT~\cite{maskCRT}, without re-training these codecs, achieving approximately 6.5\%-12.6\%  and 10.6\%-19.2\% BD-rate improvements on commonly used datasets, respectively.

\vspace{-1.0em}
\section{Related work}
\vspace{-0.5em}
\label{sec:Related_work}

Motion estimation networks play a pivotal role in learned video codecs. This is especially true for learned B-frame codecs that adopt a hierarchical prediction structure, where both small and large motion exist and must be estimated properly.

To estimate large motion, some works~\cite{accflow, tapir} generate the optical flow maps in an iterative manner. Wu \emph{et al.}~\cite{accflow} accumulate local optical flows between adjacent frames in a backward manner to estimate long-range (large) motion between two distant video frames. For the point-tracking task, Doersch \emph{et al.}~\cite{tapir} propose a two-stage approach, comprising a matching stage to compare the query point in the current frame with all the candidate points in the future frames and a refinement stage to refine the locations of the predicted points using local correlations. Despite these efforts, the need to utilize iteratively information from adjacent frames creates a long processing latency, making them computationally expensive and less suitable for our B-frame coding task. 

Other works~\cite{emdflow, film} perform motion estimation in an implicit way. Deng \emph{et al.}~\cite{emdflow} propose a Transformer-based approach that fuses the flow features in different scales and perform motion estimation. For the emerging task of near-duplicates interpolation, Reda \emph{et al.}~\cite{film} introduce a multi-scale feature extractor with shared network weights across the scales to estimate bi-directional flows in the presence of potentially large motion. It is motivated by the intuition that small motion at coarser scales is similar to large motion at finer scales. However, the efficiency of these data-driven approaches has largely to do with the quality of training data.

\vspace{-0.8em}
\section{Proposed Method}
\vspace{-0.5em}
\label{sec:Methodology}
\begin{figure*}[htb]
    \centerline{\includegraphics[width=1.0\textwidth]{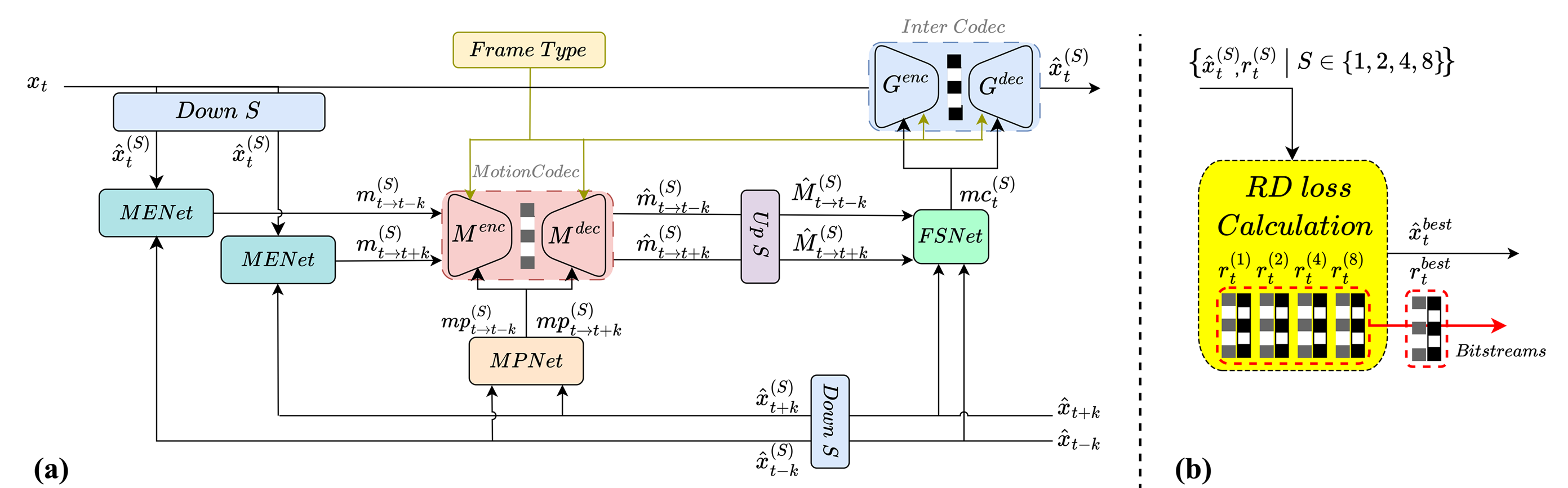}}
    \caption{Illustration of our proposed method. (a) The overall coding framework, where $x_t$ is the current coding B-frame and $\hat{x}_{t-k}, \hat{x}_{t+k}$ are the two previously reconstructed reference frames. $S$ represents the downsampling factor, which takes on 1, 2, 4, or 8. $Up$ denotes the process of upsampling to the initial resolution. (b) The online rate-distortion optimization process that  calculates the rate-distortion loss for each possible downsampling factor on a per-frame basis and chooses the bitstream with the smallest rate-distortion cost.}
    \label{fig:propose_2}
    \vspace{-1.5em}
\end{figure*}




In this paper, we propose a simple yet efficient remedy for enhancing the compression performance of pre-trained B-frame codecs that adopt hierarchical bi-directional prediction. The idea is to adapt the spatial resolution of the target and reference frames on a per-frame basis for motion estimation and coding. Our intent is to turn large motion in high-resolution video frames into small motion in their low-resolution representations to suit the capability of the motion estimation network that has been trained end-to-end on small GOPs due to dataset limitations. Our approach need not re-train the codecs, and is thus readily applicable to existing learned B-frame codecs. We validate its effectiveness with two state-of-the-art learned B-frame codecs, B-CANF~\cite{bcanf} and MaskCRT~\cite{maskCRT}. They share a similar coding architecture to Fig.~\ref{fig:propose_2}. Notably, B-CANF~\cite{bcanf} features CANF as the main compression backbone for the motion and inter-frame codecs, while MaskCRT adopts conditional Swin-Transformers.  


\vspace{-0.5em}
\subsection{System Overview}
\label{ssec:sys_overview}
As shown in Fig.~\ref{fig:propose_2}, the encoding of a B-frame $x_t \in \mathbb{R}^{3 \times W \times H}$ of width $W$ and height $H$ begins by selecting one of the downsampling factors $S$ from \{1, 2, 4, 8\}. For brevity, our following exposition and notations assume $S=2$. With this choice, the current frame $x_t$ and the reconstructed reference frames $\hat{x}_{t-k},\hat{x}_{t+k} \in \mathbb{R}^{3 \times W \times H}$ are downsampled both vertically and horizontally by a factor of 2 as $x^{(2)}_t, \hat{x}^{(2)}_{t-k}, \hat{x}^{(2)}_{t+k} \in \mathbb{R}^{3 \times \frac{W}{2} \times \frac{H}{2}}$, respectively. We then invoke the motion estimation network (MENet), which adopts SPyNet~\cite{spy}, to estimate two low-resolution optical flow maps $m^{(2)}_{t\to{t}-k}, m^{(2)}_{t\to{t}+k} \in \mathbb{R}^{2 \times \frac{W}{2} \times \frac{H}{2}}$ for $x^{(2)}_t$ with respect to $\hat{x}^{(2)}_{t-k}$ and $\hat{x}^{(2)}_{t+k}$. These generated flow maps are then collectively compressed by the conditional motion codec ($M^{enc}, M^{dec}$), where the flow map predictors $mp^{(2)}_{t\to{t}-k}, mp^{(2)}_{t\to{t}+k} \in \mathbb{R}^{2 \times \frac{W}{2} \times \frac{H}{2}}$ are generated as the conditioning signals by the motion prediction network (MPNet), which takes $\hat{x}^{(2)}_{t-k},\hat{x}^{(2)}_{t+k}$ as inputs. In other words, the low-resolution flow maps are encoded into the bitstream. To perform motion compensation, the reconstructed flow maps $\hat{m}^{(2)}_{t\to{t}-k}, \hat{m}^{(2)}_{t\to{t}+k} \in \mathbb{R}^{2 \times \frac{W}{2} \times \frac{H}{2}}$ are upsampled to the original resolution. In the present case, we upsample $\hat{m}^{(2)}_{t\to{t}-k}, \hat{m}^{(2)}_{t\to{t}+k}$ by a factor of 2, arriving at the reconstructed flow maps $\hat{M}^{(2)}_{t\to{t}-k}, \hat{M}^{(2)}_{t\to{t}+k} \in \mathbb{R}^{2 \times W \times H}$. They are then input to the frame synthesis network (FSNet), which has a structure similar to GridNet~\cite{gridnet}, together with the two high-resolution reference frames $\hat{x}_{t-k}$, $\hat{x}_{t+k}$ to perform multi-scale motion compensation in both the feature and pixel domains. The output is the temporal predictor $mc^{(2)}_t \in \mathbb{R}^{3 \times W \times H}$, which serves as the conditioning signal for the inter-frame codec ($G^{enc}, G^{dec}$) to encode $x_t$. Notably, in line with~\cite{bcanf, maskCRT}, both our motion and inter-frame codec utilize a frame-type signal, indicating the reference type (reference vs. non-reference) of each B-frame for better bit allocation.



\vspace{-1.0em}
\subsection{Online Motion Resolution Adaptation (OMRA)}
\label{sssec:select_downsampling_factor}

We conduct an exhaustive search of the best downsampling factor $S \in \{1, 2, 4, 8 \}$ for each B-frame by minimizing the per-frame rate-distortion cost:
\begin{equation}
\label{equ:rd_loss_P}
 L^{(S)} = {\lambda \cdot D(x_t, \hat{x}^{(S)}_t) + r^{(S)}_t},
\end{equation} 
where $D$ is the distortion of the reconstructed frame $x^{(S)}_t$ measured in mean-square error in the RGB domain and $r^{(S)}_t$ is the total bitrate. We follow the same $\lambda$ values in MaskCRT~\cite{maskCRT} and B-CANF~\cite{bcanf}. At inference time, we perform encoding and decoding of a B-frame with these downsampling factors and choose the one that minimizes the rate-distortion cost $L^{(S)}$. The best downsampling factor and its encoding result are signaled in the bitstream. This rate-distortion optimization process is repeated until all the B-frames are coded.

\vspace{-0.5em}
\subsection{OMRA Variants}
\label{ssec:Variants_description}

To justify our design choice, we investigate two variants of OMRA. They differ from OMRA in the way the low-resolution optical flow maps are coded and utilized. We denote OMRA as (\textit{Compress $\rightarrow$ Up\_Flow $\rightarrow$ Warp}). Recall that we first perform motion estimation in low resolution and compress the low-resolution flow maps into the bitstream, followed by upsampling the decoded flow maps to the initial resolution for temporal warping (i.e. motion compensation). The first variant (Variant A), denoted by (\textit{Up\_Flow $\rightarrow$ Compress $\rightarrow$ Warp}), upsamples the low-resolution flow maps to the initial resolution before they are compressed into the bitstream and decoded for warping. This variant is intended to recover as much motion detail as possible before motion coding. The second variant (Variant B), denoted by (\textit{Compress $\rightarrow$ Warp $\rightarrow$ Up\_MC}), compresses the low-resolution flow maps, performs warping in low resolution (i.e.~with respect to $\hat{x}^{(2)}_{t-k}, \hat{x}^{(2)}_{t+k}$), and upsamples the motion-compensated temporal predictor. It differs from the previous two approaches in that the upsamping is conducted in the intensity rather than flow domain. Section~\ref{sssec:three_implementation} compares the compression performance of these variants based on MaskCRT~\cite{maskCRT}. 

\begin{figure*}[t!]
    \begin{center}
    \begin{subfigure}{0.32\linewidth}
        \centering
        \includegraphics[width=\linewidth]{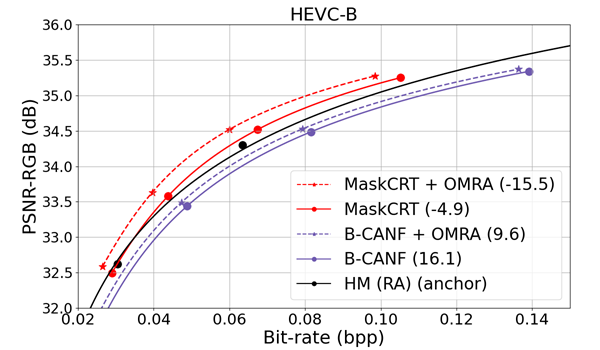}
        \caption{\centering HEVC Class B}
        \label{fig:hevcb}
    \end{subfigure}
    \begin{subfigure}{0.32\linewidth}
        \centering
        \includegraphics[width=\linewidth]{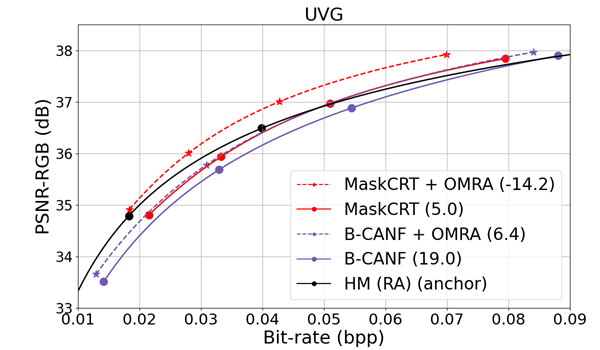}
        \caption{\centering UVG}
        \label{fig:uvg}
    \end{subfigure}
    \begin{subfigure}{0.32\linewidth}
        \centering
        \includegraphics[width=\linewidth]{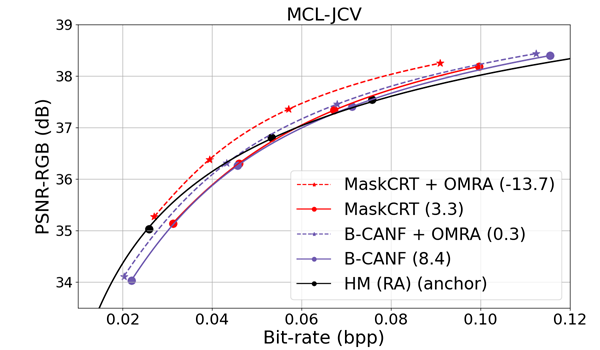}
        \caption{\centering MCL-JCV}
        \label{fig:mcl}
    \end{subfigure}

    \caption{Rate-distortion performance comparison. The intra-period is set to 32, and the values within the parentheses are BD-rate with HM-16.25 (random access)~\cite{hm16_25} serving as the anchor.}
    \label{fig:RD}
    \end{center}
    \vspace{-1.5em}
\end{figure*}

\vspace{-0.5em}
\section{Experimental Results}
\vspace{-0.5em}
\label{sec:results}
\subsection{Settings and Baselines}
\label{sec:Evaluation_methodologies}

We evaluate our method on two pre-trained state-of-the-art learned B-frame codecs, B-CANF~\cite{bcanf} and MaskCRT~\cite{maskCRT}. Note that these codecs are trained on Vimeo-90k~\cite{vimeo} with a GOP size of 5. We stress that OMRA--namely, \textit{(Compress $\rightarrow$ Up\_Flow $\rightarrow$ Warp)} in Section~\ref{sec:Methodology}--need not re-train these models. Instead, we perform rate-distortion optimization at test time to choose the downsampling factor for motion estimation and coding. We follow \cite{bcanf, maskCRT} to evaluate the compression performance on UVG~\cite{uvg}, HEVC Class B~\cite{HEVC-B}, and MCL-JCV~\cite{mcl} datasets. For each video sequence, we encode 97 frames with the intra-period set to 32. That is, the GOP size at test time is 32, as opposed to 5 for training. We measure the compression efficiency by BD-rate~\cite{bdrate} in terms of peak-signal-to-noise ratio (PSNR) in the RGB domain and bits-per-pixel (bpp). Negative and positive BD-rate numbers suggest rate reduction and inflation, respectively. The baseline methods for comparison are B-CANF~\cite{bcanf} and MaskCRT~\cite{maskCRT} without our OMRA. The anchor is HM-16.25~\cite{hm16_25} with the configuration \emph{encoder\_randomaccess\_main\_rext.cfg}.


\vspace{-1.0em}
\subsection{Rate-Distortion Comparison}
\label{sec:overall_performance}

Fig.~\ref{fig:RD} presents the rate-distortion performance of the competing methods. Applying OMRA to MaskCRT~\cite{maskCRT} and B-CANF~\cite{bcanf}, denoted as MaskCRT + OMRA and B-CANF + OMRA, respectively, yields significant coding gains. As compared to MaskCRT~\cite{maskCRT}, the additional BD-rate savings achieved by OMRA are seen to be 10.6\%, 19.2\%, 17.0\% on HEVC Class B, UVG, and MCL-JCV, respectively. Likewise, we observe 6.5\%, 12.6\%, 8.1\% BD-rate improvements with B-CANF~\cite{bcanf}. Although conceptually simple, our scheme is effective on both learned B-frame codecs.  


\begin{table}[t]
\caption{\justifying The per-sequence BD-rates of our OMRA on UVG and HEVC Class B datasets. The anchor is MaskCRT~\cite{maskCRT} and B-CANF~\cite{bcanf} without OMRA, respectively.}

\centering
\small
\begin{tabular}{cccc}
\hline 
\multirow{2}{*}{{Dataset}}                          & \multicolumn{1}{c}{\multirow{2}{*}{{Sequence}}}            & \multicolumn{2}{c}{{BD-rate (\%) PSNR-RGB}}                                            \\  \cline{3-4}                                 & \multicolumn{1}{c}{}                         & \multicolumn{1}{c}{{MaskCRT}} & \multicolumn{1}{c}{{B-CANF}} \\\hline
\multirow{8}{*}{UVG}                          & \multicolumn{1}{c}{Beauty}            & \multicolumn{1}{c}{-2.90}   & \multicolumn{1}{c}{-1.38}                                                                       \\  
                                              & \multicolumn{1}{c}{Bosphorus}         & \multicolumn{1}{c}{-9.71}   & \multicolumn{1}{c}{-2.66}                                                                    \\ 
                                              & \multicolumn{1}{c}{HoneyBee}          & \multicolumn{1}{c}{-0.98}     & \multicolumn{1}{c}{0.00}                                                                  \\ 
                                              & \multicolumn{1}{c}{Jockey}            & \multicolumn{1}{c}{-48.73}     & \multicolumn{1}{c}{-32.41}                                                                  \\  
                                              & \multicolumn{1}{c}{ReadySteadyGo}     & \multicolumn{1}{c}{-27.77}    & \multicolumn{1}{c}{-20.29}                                                                   \\ 
                                              & \multicolumn{1}{c}{ShakeNDry}         & \multicolumn{1}{c}{-0.41}      & \multicolumn{1}{c}{0.00}                                                                 \\ 
                                              & \multicolumn{1}{c}{YatchRide}         & \multicolumn{1}{c}{-6.98}     & \multicolumn{1}{c}{-3.19}                                                                                                                         \\ \hline
    \multirow{6}{*}{HEVC-B}                          & \multicolumn{1}{c}{BasketballDrive}            & \multicolumn{1}{c}{-21.16}    & \multicolumn{1}{c}{-13.97}                                                                \\  
                                                  & \multicolumn{1}{c}{BQTerrace}         & \multicolumn{1}{c}{-8.18}         & \multicolumn{1}{c}{-5.48}                                                              \\ 
                                                  & \multicolumn{1}{c}{Cactus}          & \multicolumn{1}{c}{-2.91}     & \multicolumn{1}{c}{-0.06}                                                                  \\ 
                                                  & \multicolumn{1}{c}{Kimono1}            & \multicolumn{1}{c}{-12.05}     & \multicolumn{1}{c}{-6.24}                                                                  \\  
                                              & \multicolumn{1}{c}{ParkScene}     & \multicolumn{1}{c}{-2.42}              & \multicolumn{1}{c}{-0.05}                                                                   \\ \hline
\end{tabular}
\label{tab:per_seq}
\vspace{-1.0em}
\end{table}
Table~\ref{tab:per_seq} further presents the per-sequence BD-rate for UVG~\cite{uvg} and HEVC Class B~\cite{HEVC-B}. We observe that the improvements are most obvious on fast-motion sequences (e.g.~$BasketballDrive$, $ReadySteadyGo$ and $Jockey$), where the domain shift between training and test is most significant. In other words, the MENet trained on GOP 5 performs poorly when tested on GOP 32 due to large motion. In contrast, the gains are less considerable on slow-motion sequences (e.g.~$HoneyBee$ and $ParkScene$), where the MENet trained on GOP 5 is able to work well in estimating small motion even with GOP 32. In this case, performing motion estimation and coding in low resolution does not improve the quality of the temporal predictor. 


\vspace{-1.0em}
\begin{figure}[t!]

    \centering
    \includegraphics[width=\linewidth]{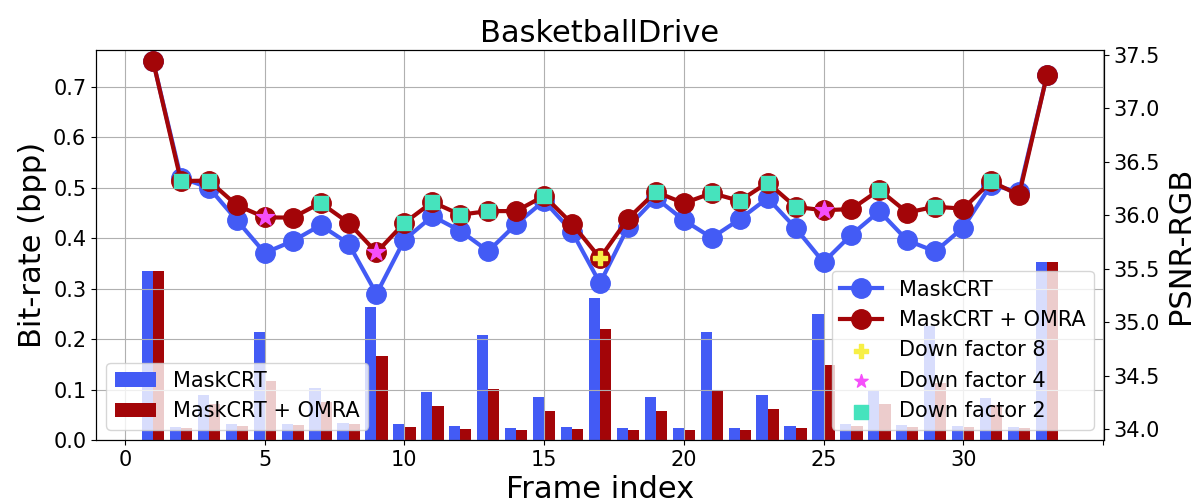}
    \caption{Profiles of the per-frame PSNR’s and the per-frame bitrate for the first GOP in $BasketballDrive$, with and without OMRA. '$\blacksquare$', '$\star$', '$+$' represent the downsampling factors of 2, 4, 8, respectively. Points without these symbols indicate no downsampling.}
    \label{fig:ba_PerFrame}
    \vspace{-1.2em}
\end{figure}

\subsection{Downsampling Factors versus Temporal Levels}
\begin{figure}[t!]
    \begin{center}
    \begin{subfigure}{\linewidth}
        \centering
        \includegraphics[width=\linewidth]{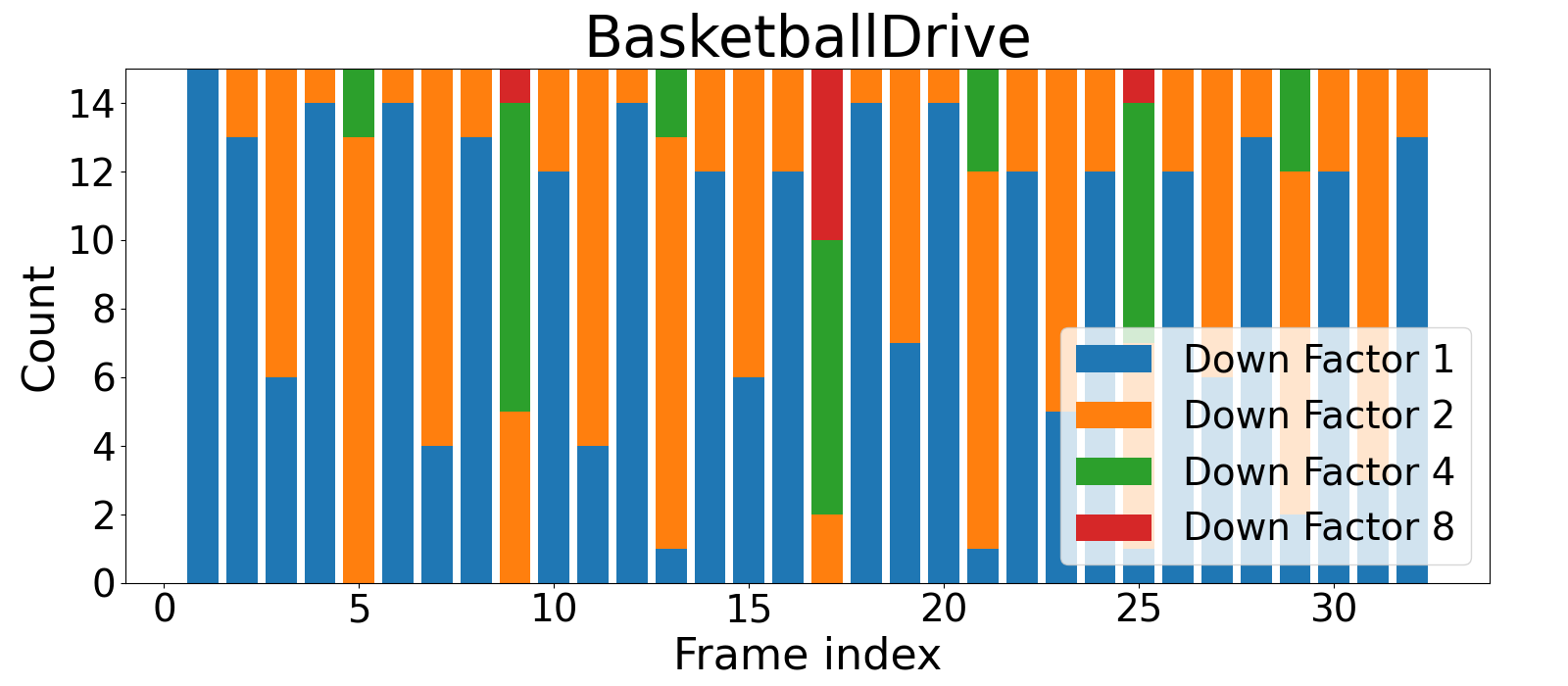}
        \label{fig:ba_scale}
    \end{subfigure}
    \begin{subfigure}{\linewidth}
    \vspace{-1em}
        \centering
        \includegraphics[width=\linewidth]{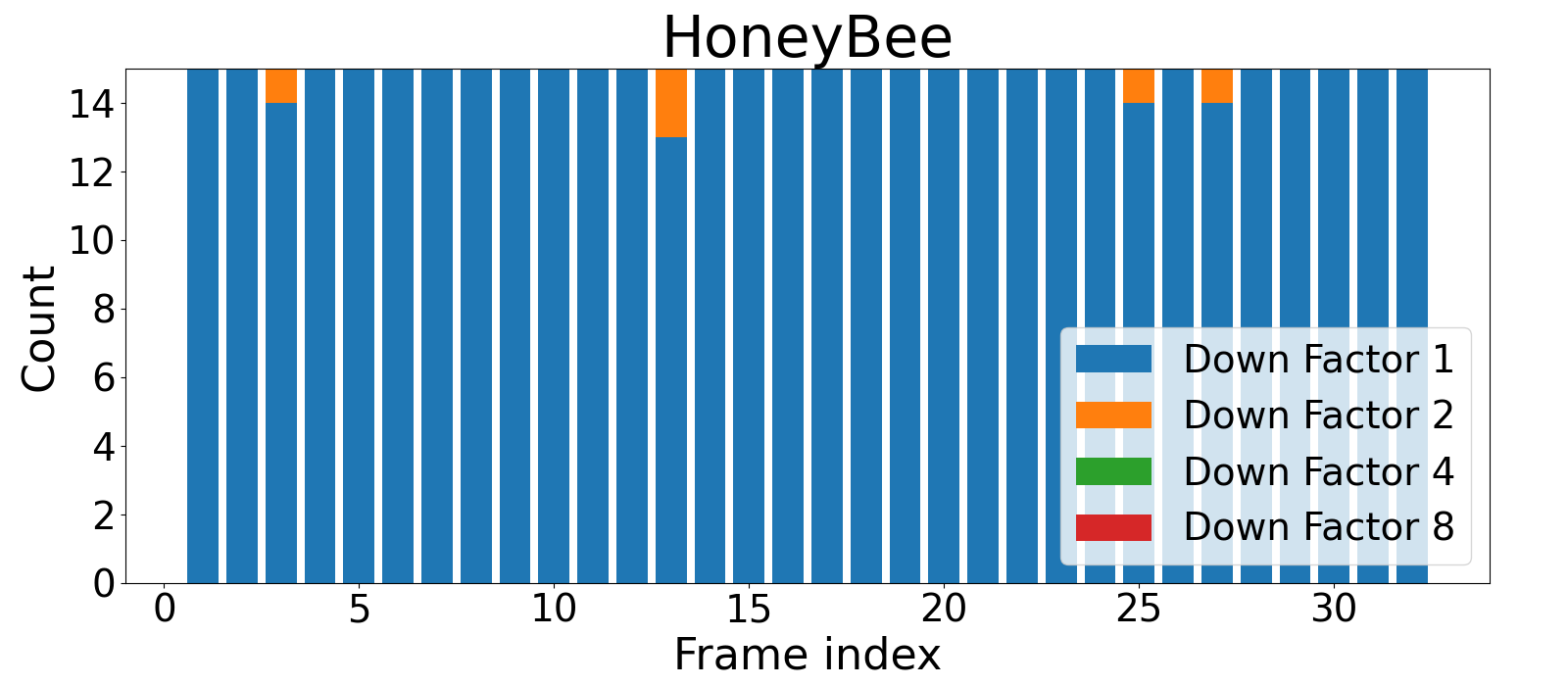}
        \label{fig:honey_scale}
    \end{subfigure}
    \vspace{-3em}
    \caption{Relative frequencies of the downsampling factors chosen for each frame in $BasketballDrive$ and $HoneyBee$.}
    \label{fig:scale_factor_count}
    \end{center}
    \vspace{-2.5em}
\end{figure}

\vspace{-0.5em}
\label{sec:overall_performance}
Fig.~\ref{fig:ba_PerFrame} visualizes the PSNR and bitrate profiles in a hierarchically predicted GOP of size 32 for $BasketballDrive$, a fast-motion sequence. We take MaskCRT as the baseline. As shown, our OMRA consistently improves the reconstruction quality of MaskCRT across video frames while reducing the required bitrates. Remarkably, the improvements are most obvious at video frames (e.g. frame 9, 17, and 25) of lower temporal levels where the longer prediction distance between the target and reference frames incurs larger motion (see Fig.~\ref{fig:coding_structure} for the definition of temporal levels). As expected, larger downsampling factors are chosen at lower temporal levels as a result of the domain shift issue. This is corroborated in a further in-depth analysis presented in Fig.~\ref{fig:scale_factor_count}, where the relative frequencies of downsampling factors chosen for each video frame are collected over 15 GOPs for $BasketballDrive$ and $HoneyBee$. It is worth noting that in the slow-motion sequence $HoneyBee$, the best strategy is not to invoke downsampling in order to retain as much motion detail as possible. In this case, the MENet trained on small GOPs is able to well estimate slow motion.  




\vspace{-1.1em}
\subsection{Subjective Quality Comparison}
\begin{figure}[t!]
\centering
\footnotesize
{
\begin{tabular}{ccc}

    \includegraphics[width=0.3\linewidth]{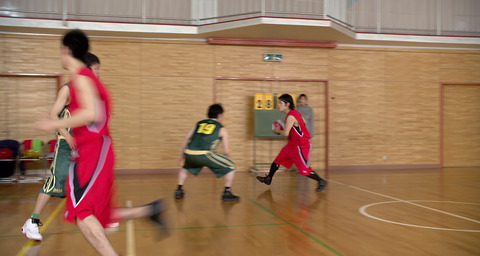} &
    \includegraphics[width=0.3\linewidth]{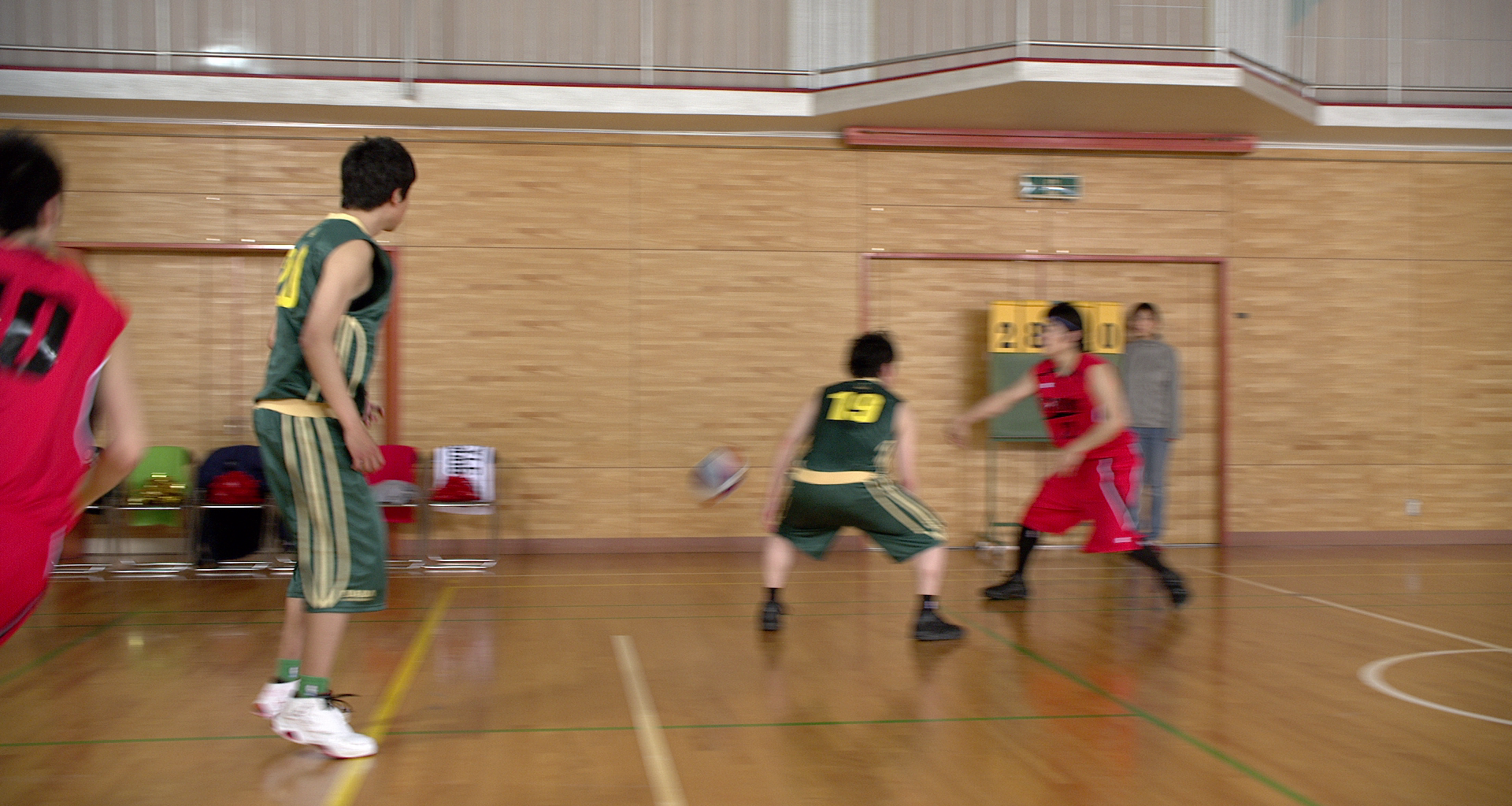} &
    \includegraphics[width=0.3\linewidth]{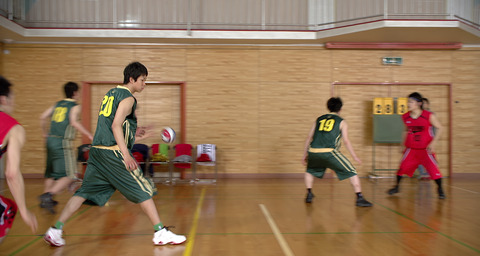} \\
     \multicolumn{1}{c}{(a) Past ($x_{1}$)} &  \multicolumn{1}{c}{(b) Target ($x_{17}$)} &  \multicolumn{1}{c}{(c) Future ($x_{33}$)}  \\

    \includegraphics[width=0.3\linewidth]{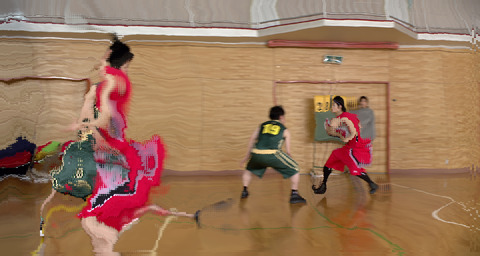} &
    \includegraphics[width=0.3\linewidth]{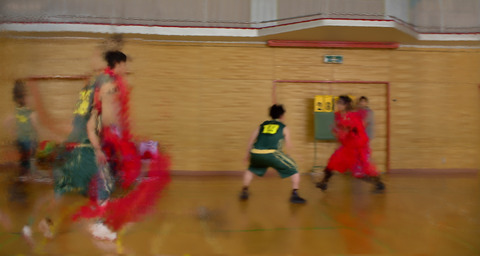}     &
    \includegraphics[width=0.3\linewidth]{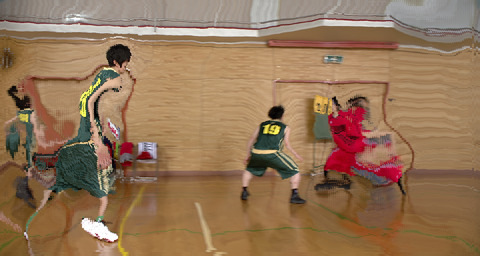}  \\
     \multicolumn{1}{c}{(d) $W(x_{1}, F_{17\to{1}})$} &  \multicolumn{1}{c}{(e) $mc^{(S)}_{17}$} &  \multicolumn{1}{c}{(f) $W(x_{33}, F_{17\to{33}})$ }  \\

     \multicolumn{1}{c}{(MaskCRT + OMRA)} &  \multicolumn{1}{c}{(MaskCRT + OMRA)} &  \multicolumn{1}{c}{(MaskCRT + OMRA)}  \\

    \includegraphics[width=0.3\linewidth]{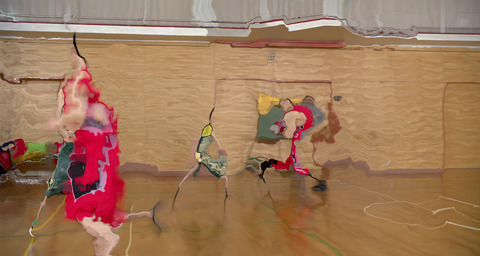} &
    \includegraphics[width=0.3\linewidth]{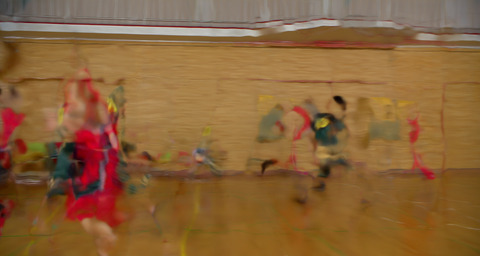}      &
    \includegraphics[width=0.3\linewidth]{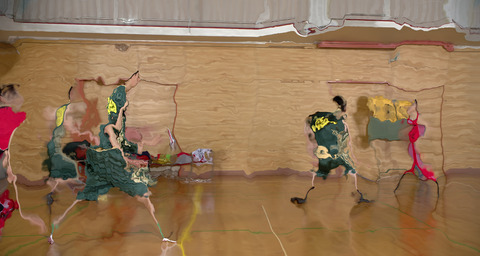} \\

     \multicolumn{1}{c}{(g) $W(x_{1}, F_{17\to{1}})$} &  \multicolumn{1}{c}{(h) $mc^{(S)}_{17}$} &  \multicolumn{1}{c}{(i) $W(x_{33}, F_{17\to{33}})$}  \\

     \multicolumn{1}{c}{(MaskCRT)} &  \multicolumn{1}{c}{(MaskCRT)} &  \multicolumn{1}{c}{(MaskCRT)}  \\
\end{tabular}
}

\vspace{-1mm}

\caption{Visualization of the warped frames and temporal predictors in $BasketballDrive$. The first row is input frames, with (a)(c) being the reference frames and (b) the target frame. The second row is results with OMRA, where (d) and (f) are temporally warped frames from (a) and (c), respectively. (e) is the temporal predictor. Likewise, the third row is the corresponding results without OMRA. $W(x_{t'},F_{t \to t'})$ denotes backward warping, where $x_{t'}$ is the reference frame to be warped and $F_{t \to t'}$ describes the backward flow from $x_t$ to $x_{t'}$.}

\label{fig:ba_visualization}
\end{figure}

\vspace{-0.5em}
\label{ssec:ablation}
Fig.~\ref{fig:ba_visualization} visualizes the warped frames and the temporal predictors for $BasketballDrive$. The second and third rows are MaskCRT with OMRA ($S=8$) and MaskCRT without OMRA, respectively. A comparison of Fig.~\ref{fig:ba_visualization}(d) and Fig.~\ref{fig:ba_visualization}(g) (respectively, Fig.~\ref{fig:ba_visualization}(f) and Fig.~\ref{fig:ba_visualization}(i)) shows that OMRA is able to generate a better-quality warped frame at test time. As a result, the final temporal predictor with OMRA in Fig.~\ref{fig:ba_visualization}(e) shows better quality than that without OMRA in Fig.~\ref{fig:ba_visualization}(h). This confirms that our OMRA is an effective means to mitigate the domain shift, particularly in fast-motion sequences where the domain shift is most obvious. 

\begin{figure}[t!]
\centering
\footnotesize
{
\begin{tabular}{ccc}

    \includegraphics[width=0.3\linewidth]{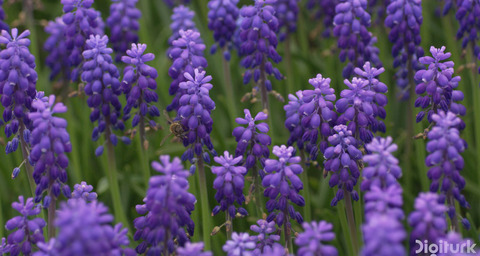} &
    \includegraphics[width=0.3\linewidth]{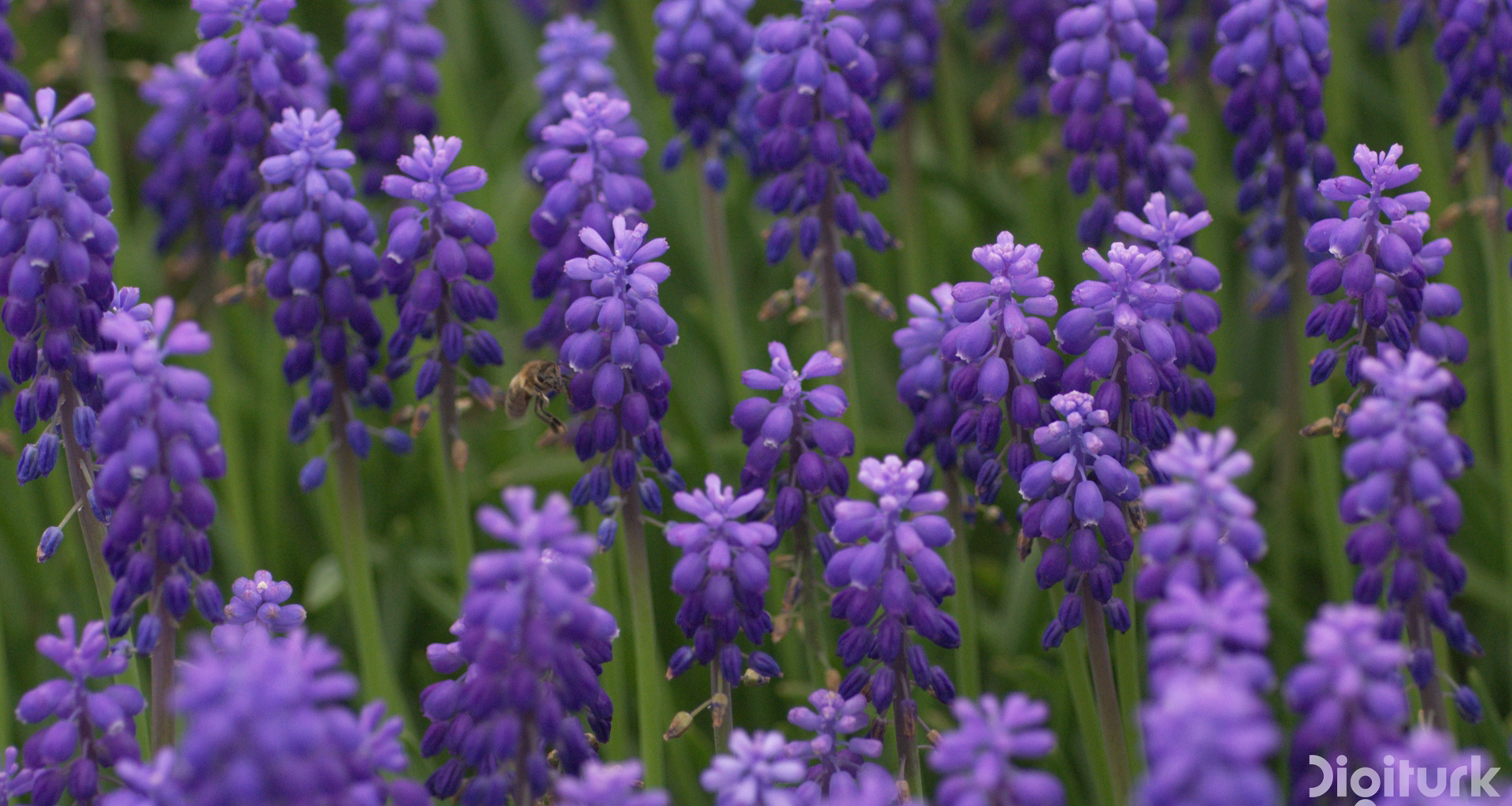} &
    \includegraphics[width=0.3\linewidth]{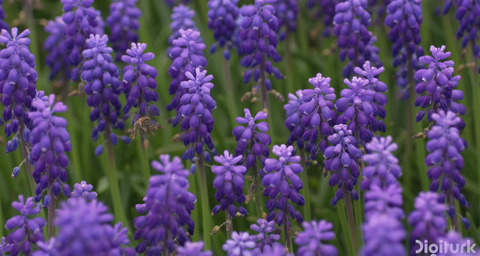} \\
     \multicolumn{1}{c}{(a) Past ($x_{1}$)} &  \multicolumn{1}{c}{(b) Target ($x_{17}$)} &  \multicolumn{1}{c}{(c) Future ($x_{33}$)}  \\

    \includegraphics[width=0.3\linewidth]{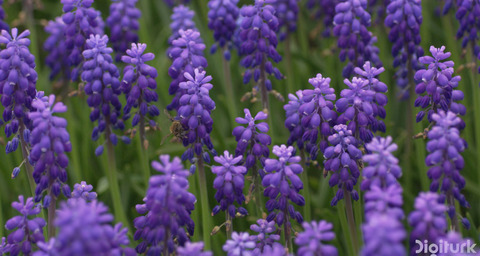} &
    \includegraphics[width=0.3\linewidth]{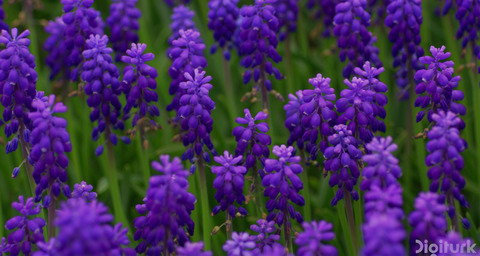}     &
    \includegraphics[width=0.3\linewidth]{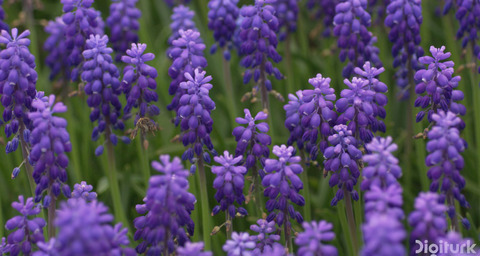}  \\
     \multicolumn{1}{c}{(d) $W(x_{1}, F_{17\to{1}})$} &  \multicolumn{1}{c}{(e) $mc^{(S)}_{17}$} &  \multicolumn{1}{c}{(f) $W(x_{33}, F_{17\to{33}})$}  \\

     \multicolumn{1}{c}{(MaskCRT + OMRA)} &  \multicolumn{1}{c}{(MaskCRT + OMRA)} &  \multicolumn{1}{c}{(MaskCRT + OMRA)}  \\

    \includegraphics[width=0.3\linewidth]{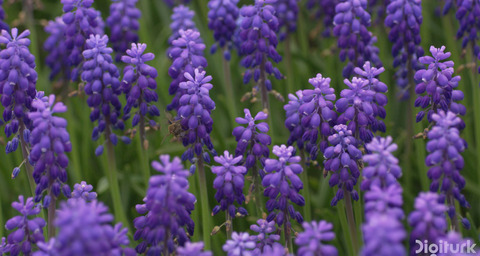} &
    \includegraphics[width=0.3\linewidth]{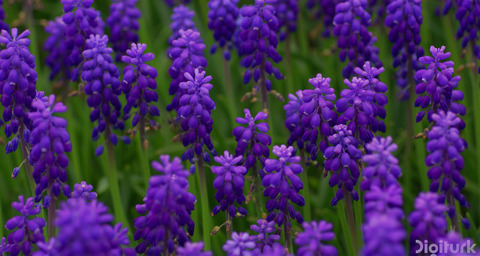}      &
    \includegraphics[width=0.3\linewidth]{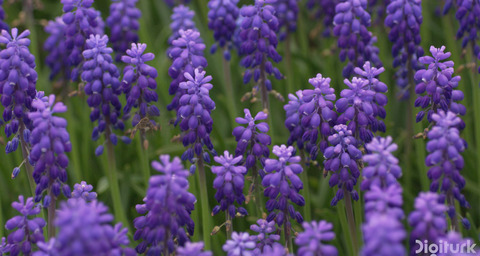} \\

     \multicolumn{1}{c}{(g) $W(x_{1}, F_{17\to{1}})$} &  \multicolumn{1}{c}{(h) $mc^{(S)}_{17}$} &  \multicolumn{1}{c}{(i) $W(x_{33}, F_{17\to{33}})$}  \\

     \multicolumn{1}{c}{(MaskCRT)} &  \multicolumn{1}{c}{(MaskCRT)} &  \multicolumn{1}{c}{(MaskCRT)}  \\
\end{tabular}
}

\vspace{-1mm}

\caption{Visualization of the warped frames and the temporal predictors for $HoneyBee$. All the notations follow the caption in Fig.~\ref{fig:ba_visualization}.}

    \label{fig:honey_visualization}
\end{figure}

Fig.~\ref{fig:honey_visualization} visualizes the results for $HoneyBee$, a slow-motion sequence where the motion estimation network trained on GOP 5 is still able to predict well small motion on GOP 32. In the present case, OMRA does not bring any noticeable gain in terms of the subjective results. 

Last but not least, the color shift between the temporal predictors (parts (e)(h) in Figs.~\ref{fig:ba_visualization} and~\ref{fig:honey_visualization}) and the ground-truths (part (b) in Figs.~\ref{fig:ba_visualization} and~\ref{fig:honey_visualization}) has to do with the fact that the temporal predictor in our framework is a latent signal with no regularization.  



\subsection{Ablation Experiments}
\label{ssec:ablation}


\begin{table}[!t]
    \centering
    \footnotesize
    \caption{BD-rate comparison with intra-period 32. The anchor is HM-16.25 (random access)~\cite{hm16_25}. 
}
    \label{tab:exp_BD_PSNR}
    \begin{tabular}{cccc}
        \hline
        \multirow{2}{*}{Method}  & \multicolumn{3}{c}{\multirow{1}{*}{BD-rate (\%) PSNR-RGB}}\\
            
            \cline{2-4}
            
            &\multirow{1}{*}{HEVC-B}   & \multirow{1}{*}{UVG}  & \multirow{1}{*}{MCL-JCV}  \\
                \hline
        
        MaskCRT\_SPy            & -4.9 & 5.0 & 3.3  \\
        MaskCRT\_SPy + OMRA  & \textbf{-15.5} & \textbf{-14.2} & \textbf{-13.7}  \\
        MaskCRT\_PWC            & -0.2 & -1.8 & 3.1  \\
        MaskCRT\_PWC + OMRA  & \textbf{-9.9} & \textbf{-7.9}  & \textbf{-8.9}  \\
        \hline


    \end{tabular}
\end{table}

\subsubsection{Effect of Motion Estimation Networks}
To validate that our OMRA is agnostic to the motion estimation network, we replace SPyNet~\cite{spy} in MaskCRT~\cite{maskCRT} with PWCNet~\cite{pwc}. The updated model is re-trained end-to-end on Vimeo-90k~\cite{vimeo} with a small GOP size of 5. At test time, we apply OMRA with GOP 32. From Table~\ref{tab:exp_BD_PSNR}, our OMRA is still effective with PWCNet~\cite{pwc}, showing 9.7\%, 6.1\%, 12.0\% BD-rate improvements on HEVC Class B, UVG, MCL-JCV, respectively, as compared to MaskCRT\_PWC without OMRA. Note that PWCNet~\cite{pwc} is heavier and more capable than SPyNet~\cite{spy}. This explains the marginally reduced gain of OMRA with PWCNet~\cite{pwc}.

\vspace{-1em}
\begin{table}[t]
    \centering
    \caption[]{BD-rate comparison of OMRA and its variants. The anchor for BD-rate evaluation is MaskCRT~\cite{maskCRT}.}
    \scriptsize
    
    \begin{tabular}{ccccc}
        \hline
        \multicolumn{5}{c}{\multirow{1}{*}{BD-rate (\%) PSNR-RGB}} \\
        \hline
         \multicolumn{1}{c}{\multirow{1}{*}{Settings}} & \multicolumn{1}{c}{\multirow{1}{*}{Schemes}} &  \multicolumn{1}{c}{\multirow{1}{*}{HEVC-B}}& 
        \multicolumn{1}{c}{\multirow{1}{*}{UVG}} &
        \multicolumn{1}{c}{\multirow{1}{*}{MCL-JCV}}  \\
        
        \hline
          \multicolumn{1}{c}{OMRA} & \multicolumn{1}{c}{Compress $\rightarrow$ Up\_Flow $\rightarrow$ Warp} &   \multicolumn{1}{c}{-11.0}   &   \multicolumn{1}{c}{-17.7}     &   \multicolumn{1}{c}{-15.9}   \\ 
        
           \multicolumn{1}{c}{Variant A} & \multicolumn{1}{c}{Up\_Flow $\rightarrow$ Compress $\rightarrow$ Warp}   &   \multicolumn{1}{c}{-1.9}    &   \multicolumn{1}{c}{-5.0}   &   \multicolumn{1}{c}{-2.0}  \\
          \multicolumn{1}{c}{Variant B} & \multicolumn{1}{c}{Compress $\rightarrow$ Warp $\rightarrow$ Up\_MC}&   \multicolumn{1}{c}{0.0}    &   \multicolumn{1}{c}{0.0}   &   \multicolumn{1}{c}{0.0}\\
            \hline

    \end{tabular}
    \label{tab:three_implementation}
    \vspace{-1.5em}
\end{table}
\subsubsection{OMRA versus Its Variants}
\label{sssec:three_implementation}
Table~\ref{tab:three_implementation} presents BD-rate results for various OMRA variants when implemented with MaskCRT~\cite{maskCRT}, which serves as the anchor for BD-rate evaluation. We see that OMRA (\textit{Compress $\rightarrow$ Up\_Flow $\rightarrow$ Warp}) achieves the highest BD-rate saving as compared to its two other variants. 

Variant A (\textit{Up\_Flow $\rightarrow$ Compress $\rightarrow$ Warp}) marginally enhances BD-rate savings, but is not as effective as OMRA. This suggests that the upsampled flow maps, although retaining more motion detail, need more bits to represent and are subject to compression artifacts. The coded flow maps are not so effective for warping in the rate-distortion sense. 

Variant B (\textit{Compress $\rightarrow$ Warp $\rightarrow$ Up\_MC}) shows no improvement, compared to the anchor. As compared to OMRA, which upsamples the coded flow maps and then performs temporal warping in high resolution, this variant performs temporal warping in low resolution and subsequently upsamples the resulting low-resolution video frame. Generally, upsampling a video frame is much more challenging than upsampling a flow map. The intensity data have much more high-frequency components to recover. As such, Variant B is not as effective as OMRA. Moreover, the simple bilinear interpolation is not very effective. As such, Variant B does not have any competitive advantage over the anchor. These results justify the design of OMRA. 

\begin{table}[t]
\centering
    \caption{Encoding complexity comparison.}
    \label{tab:complexity}
    \begin{tabular}{ccc}
    \hline
    Method & Encoding Time & Encoding MACs \\ \hline
    MaskCRT &  1.93s & 2.03 M/pixel  \\
    MaskCRT + OMRA   &  6.48s & 5.97 M/pixel  \\ \hline
    B-CANF   &  1.52s & 3.08 M/pixel \\
    B-CANF + OMRA   &  3.38s & 8.20 M/pixel \\ \hline
    \end{tabular}
\end{table}




\subsubsection{Encoding Complexity}


Table~\ref{tab:complexity} presents the impact of OMRA on the encoding complexity in terms of the encoding runtime and the number of multiply–accumulate operations (MACs) per pixel. The encoding times are averaged over the first 93 coded B-frames of $Beauty$ sequence in UVG dataset with GOP 32. The reported numbers in Table~\ref{tab:complexity} are collected on NVIDIA Tesla V100.

Both B-CANF~\cite{bcanf} and MaskCRT~\cite{maskCRT} exhibit increased complexity after applying OMRA. This increase is attributed to the fact that we conduct an exhaustive search to minimize the per-frame rate-distortion cost during encoding. We acknowledge that OMRA leads to higher encoding runtimes and complexity. How to predict the downsampling factor on a per-frame basis without an exhaustive search is among our future work. Notably, OMRA has a negligible effect on the decoding complexity, although requiring additional bilinear interpolation with respect to the coded flow maps. 





\vspace{-1.0em}
\section{Conclusion}
\vspace{-1.0em}
\label{sec:conclusion}

In this paper, we present an online motion resolution adaptation method (OMRA) to address the domain shift issue for learned B-frame codecs that adopt hierarchical bi-directional prediction. Our findings include (1) adapting the spatial resolution of input video frames is an effective means to turn large motion into small motion in order to suit the capability of the motion estimation network in a pre-trained B-frame codec, (2) the optimal spatial resolution needs to be searched on a per-frame basis, and (3) the downsampling/upsampling is more preferred in the flow rather than intensity domain. OMRA requires no re-training of the codec and brings considerable coding gains. However, how to avoid an exhaustive search of the optimal spatial resolution is among our future work. 
\vspace{-1.0em}


\bibliographystyle{IEEEbib}
\bibliography{strings,refs}

\begin{thebibliography}{10}

\bibitem{dvclu}
Guo Lu, Wanli Ouyang, Dong Xu, Xiaoyun Zhang, Chunlei Cai, and Zhiyong Gao,
\newblock ``Dvc: An end-to-end deep video compression framework,''
\newblock in {\em Proceedings of the IEEE/CVF Conference on Computer Vision and Pattern Recognition}, 2019, pp. 11006--11015.

\bibitem{ssf}
Eirikur Agustsson, David Minnen, Nick Johnston, Johannes Balle, Sung~Jin Hwang, and George Toderici,
\newblock ``Scale-space flow for end-to-end optimized video compression,''
\newblock in {\em Proceedings of the IEEE/CVF Conference on Computer Vision and Pattern Recognition}, 2020, pp. 8503--8512.

\bibitem{nvc}
Haojie Liu, Ming Lu, Zhan Ma, Fan Wang, Zhihuang Xie, Xun Cao, and Yao Wang,
\newblock ``Neural video coding using multiscale motion compensation and spatiotemporal context model,''
\newblock {\em IEEE Transactions on Circuits and Systems for Video Technology}, 2020.

\bibitem{fvc}
Zhihao Hu, Guo Lu, and Dong Xu,
\newblock ``Fvc: A new framework towards deep video compression in feature space,''
\newblock in {\em Proceedings of the IEEE/CVF Conference on Computer Vision and Pattern Recognition}, 2021, pp. 1502--1511.

\bibitem{Wu_2018}
Chao-Yuan Wu, Nayan Singhal, and Philipp Krahenbuhl,
\newblock ``Video compression through image interpolation,''
\newblock in {\em Proceedings of the European conference on computer vision (ECCV)}, 2018, pp. 416--431.

\bibitem{hlvc}
Ren Yang, Fabian Mentzer, Luc~Van Gool, and Radu Timofte,
\newblock ``Learning for video compression with hierarchical quality and recurrent enhancement,''
\newblock in {\em Proceedings of the IEEE/CVF Conference on Computer Vision and Pattern Recognition}, 2020, pp. 6628--6637.

\bibitem{murat_lhbdc}
M.~Akın Yılmaz and A.~Murat Tekalp,
\newblock ``End-to-end rate-distortion optimized learned hierarchical bi-directional video compression,''
\newblock {\em IEEE Transactions on Image Processing}, vol. 31, pp. 974--983, 2022.

\bibitem{BEPIC}
Reza Pourreza and Taco Cohen,
\newblock ``Extending neural p-frame codecs for b-frame coding,''
\newblock {\em 2021 IEEE/CVF International Conference on Computer Vision (ICCV)}, pp. 6660--6669, 2021.

\bibitem{bcanf}
Mu-Jung Chen, Yi-Hsin Chen, and Wen-Hsiao Peng,
\newblock ``B-canf: Adaptive b-frame coding with conditional augmented normalizing flows,''
\newblock {\em IEEE Transactions on Circuits and Systems for Video Technology}, pp. 1--1, 2023.

\bibitem{maskCRT}
Yi-Hsin Chen, Hong-Sheng Xie, Cheng-Wei Chen, Zong-Lin Gao, Wen-Hsiao Peng, Martin Benjak, and Jörn Ostermann,
\newblock ``Maskcrt: Masked conditional residual transformer for learned video compression,''
\newblock {\em arXiv preprint arXiv:2312.15829}, 2023.

\bibitem{TLZMC}
David Alexandre, Hsueh-Ming Hang, and Wen-Hsiao Peng,
\newblock ``Hierarchical b-frame video coding using two-layer canf without motion coding,''
\newblock in {\em Proceedings of the IEEE/CVF Conference on Computer Vision and Pattern Recognition (CVPR)}, June 2023, pp. 10249--10258.

\bibitem{vimeo}
Tianfan Xue, Baian Chen, Jiajun Wu, Donglai Wei, and William~T Freeman,
\newblock ``Video enhancement with task-oriented flow,''
\newblock {\em International Journal of Computer Vision}, vol. 127, no. 8, pp. 1106--1125, 2019.

\bibitem{accflow}
Guangyang Wu, Xiaohong Liu, Kunming Luo, Xi~Liu, Qingqing Zheng, Shuaicheng Liu, Xinyang Jiang, Guangtao Zhai, and Wenyi Wang,
\newblock ``Accflow: backward accumulation for long-range optical flow,''
\newblock in {\em Proceedings of the IEEE/CVF International Conference on Computer Vision}, 2023, pp. 12119--12128.

\bibitem{tapir}
Carl Doersch, Yi~Yang, Mel Vecerik, Dilara Gokay, Ankush Gupta, Yusuf Aytar, Joao Carreira, and Andrew Zisserman,
\newblock ``Tapir: Tracking any point with per-frame initialization and temporal refinement,''
\newblock in {\em Proceedings of the {IEEE/CVF} International Conference on Computer Vision ({ICCV})}, 2023, pp. 10061--10072.

\bibitem{emdflow}
Changxing Deng, Ao~Luo, Haibin Huang, Shaodan Ma, Jiangyu Liu, and Shuaicheng Liu,
\newblock ``Explicit motion disentangling for efficient optical flow estimation,''
\newblock in {\em Proceedings of the IEEE/CVF International Conference on Computer Vision}, 2023, pp. 9521--9530.

\bibitem{film}
Fitsum Reda, Janne Kontkanen, Eric Tabellion, Deqing Sun, Caroline Pantofaru, and Brian Curless,
\newblock ``Film: Frame interpolation for large motion,''
\newblock in {\em European Conference on Computer Vision (ECCV)}, 2022.

\bibitem{spy}
Anurag Ranjan and Michael~J Black,
\newblock ``Optical flow estimation using a spatial pyramid network,''
\newblock in {\em Proceedings of the IEEE conference on computer vision and pattern recognition}, 2017, pp. 4161--4170.

\bibitem{gridnet}
Damien Fourure, R{\'e}mi Emonet, Elisa Fromont, Damien Muselet, Alain Tr{\'e}meau, and Christian Wolf,
\newblock ``Residual conv-deconv grid network for semantic segmentation,''
\newblock in {\em Proceedings of the British Machine Vision Conference, 2017}, 2017.

\bibitem{hm16_25}
``Hm-16.25,'' https://vcgit.hhi.fraunhofer.de/jvet/HM/,
\newblock Accessed: 2023-10-30.

\bibitem{uvg}
Alexandre Mercat, Marko Viitanen, and Jarno Vanne,
\newblock ``Uvg dataset: 50/120fps 4k sequences for video codec analysis and development,''
\newblock in {\em Proceedings of the 11th ACM Multimedia Systems Conference}, 2020, pp. 297--302.

\bibitem{HEVC-B}
F.~Bossen et~al.,
\newblock ``Common test conditions and software reference configurations,''
\newblock {\em JCTVC-L1100}, vol. 12, no. 1, 2013.

\bibitem{mcl}
Haiqiang Wang, Weihao Gan, Sudeng Hu, Joe~Yuchieh Lin, Lina Jin, Longguang Song, Ping Wang, Ioannis Katsavounidis, Anne Aaron, and C-C~Jay Kuo,
\newblock ``Mcl-jcv: a jnd-based h. 264/avc video quality assessment dataset,''
\newblock in {\em 2016 IEEE International Conference on Image Processing (ICIP)}. IEEE, 2016, pp. 1509--1513.

\bibitem{bdrate}
Gisle Bjontegaard,
\newblock ``Calculation of average psnr differences between rd-curves,''
\newblock {\em ITU SG16 Doc. VCEG-M33}, 2001.

\bibitem{pwc}
Deqing Sun, Xiaodong Yang, Ming-Yu Liu, and Jan Kautz,
\newblock ``Pwc-net: Cnns for optical flow using pyramid, warping, and cost volume,''
\newblock in {\em Proceedings of the IEEE conference on computer vision and pattern recognition}, 2018, pp. 8934--8943.

\end{thebibliography}

\end{document}